\documentclass[pra,a4paper,groupedaddress]{revtex4}
\usepackage{graphicx}
\usepackage{tabularx}
\usepackage{amsmath}
\usepackage{amsfonts}
\usepackage{amssymb}

\begin{document}
\title{Optical Response of Nanostructured Surfaces:\\Experimental Investigation of the Composite Diffracted Evanescent Wave Model}
\date{\today}
\author{G. Gay}\author{O. Alloschery}\author{B. Viaris de Lesegno}\altaffiliation[Present address:]{ Laboratoire Aim{\'e} Cotton, Campus
d'Orsay, 91405 Orsay, France}
\author{C. O'Dwyer}\altaffiliation[Present address:]{ Photonic Nanostructures Group
Tyndall National Institute, Lee Maltings, Cork, IRELAND}
\author{J. Weiner}
\affiliation{IRSAMC/LCAR, Universit\'e Paul Sabatier, 118 route de
Narbonne, 31062 Toulouse, France} \author{H. J. Lezec}
\affiliation{Thomas J. Watson Laboratories of Applied Physics,
California Institute of Technology, Pasadena, California 91125
USA}\affiliation{Centre National de la Recherche Scientifique, 3,
rue Michel-Ange, 75794 Paris cedex 16, France}

\begin{abstract}{Investigations of the optical response of subwavelength structure
arrays milled into thin metal films has revealed surprising
phenomena including reports of unexpectedly high transmission of
light. Many studies have interpreted the optical coupling to the
surface in terms of the resonant excitation of surface plasmon
polaritons (SPPs), but other approaches involving composite
diffraction of surface evanescent waves (CDEW) have also been
proposed. We present here a series of measurements on very simple
one-dimensional (1-D) subwavelength structures with the aim of
testing key properties of the surface waves and comparing them to
the CDEW and SPP models.}
\end{abstract}
\maketitle
\section{Introduction}
Initial reports of dramatically enhanced transmission through
arrays of subwavelength holes in thin films and
membranes\,\cite{ELG98,TPL01,GTG98} have focused attention on the
physics underlying this surprising optical response. Since the
early experiments were carried out on metal films, surface plasmon
polaritons\,\cite{Raether88,BDE03} were invoked to explain the
anomalously high transmission and to suggest new types of photonic
devices\,\cite{BDE03}. Other interpretations based on ``dynamical
diffraction" in periodic slit and hole arrays\,\cite{T99,T02} or
various kinds of resonant cavity modes in 1-D slits and slit
arrays\,\cite{CL02,VLE03} have also been proposed.  Reassessment
of the earlier data by new numerical studies\,\cite{CGS05} and new
measurements\,\cite{LT04} have prompted a sharp downward revision
of the enhanced transmission factor from $\simeq 1000$ to $\simeq
10$ and have motivated the development of a new model of surface
wave excitation termed the composite diffracted evanescent wave
(CDEW) model\,\cite{LT04}. This model builds a composite surface
wave from the large distribution of diffracted evanescent modes
(the inhomogeneous modes of the ``angular spectrum representation"
of wave fields\,\cite{MW95}) generated by a subwavelength feature
such as a hole, slit, or groove when subjected to an external
source of propagating wave excitation.  The CDEW model predicts
three specific surface wave properties.  First, the surface wave
is a composite or ``wave packet" of modes each evanescent in the
direction normal to the surface.  The surface wave packet exhibits
well-defined nodal positions spaced by a characteristic
wavelength, $\lambda_{\mathrm{surf}}$; second, the appearance of
the first node at a distance of $\lambda_{\mathrm{surf}}/2$ from
the subwavelength launch site (essentially a phase delay of
$\pi/2$ with respect to the E-field of the external driving
source); and third, an amplitude decreasing inversely with
distance from the launch site. We present here the results of a
series of experiments on very simple 1-D subwavelength surface
structures designed to investigate these predictions and thus
assess the validity of the model.

\section{Summary of the CDEW Model}
\begin{figure}\centering
\includegraphics[width=0.5\columnwidth]{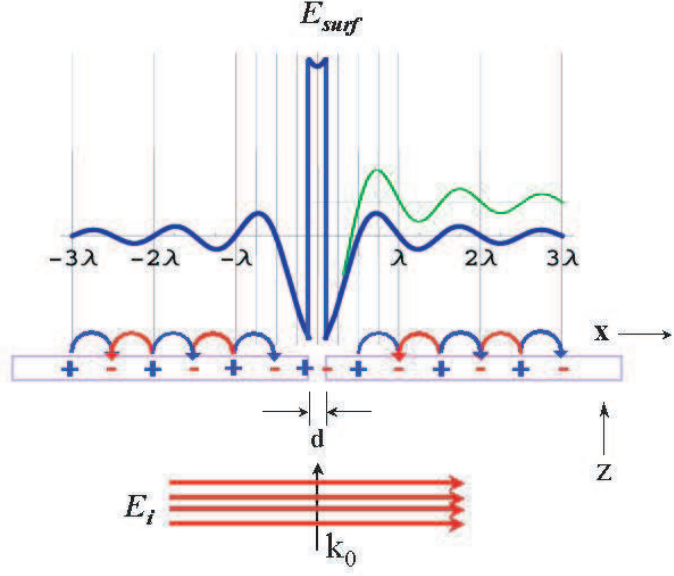}\caption{Essential elements of the CDEW model.
The incoming plane wave $E_i$ with $k_0=2\pi/\lambda_0$ in air
($n=1$) is linearly polarised parallel to the plane of the
structure and perpendicular to the slit of subwavelength width
$d$.  A fraction of the incoming light $E_{\mathrm{surf}}$ forms
the composite diffracted wave in the $\pm x$ directions, and the
blue trace (displaced above the surface for clarity) shows
$E_{\mathrm{surf}}$.  The alternating blue and red loops indicate
the field lines induced by the CDEW near the surface of the silver
film. The green trace (offset above the blue trace for clarity)
shows the cosine representation of the CDEW expressed by
Eq.\,\ref{Eq:CDEWcosrep} and closely approximating
Eq.\,\ref{Eq:Kowarz expression-a} for $|x|\geq 3/4\,\lambda$.  The
wavelength $\lambda_{\mathrm{surf}}=\lambda_0/n_{\mathrm{surf}}$
where $n_{\mathrm{surf}}$ is the surface index of refraction.}
 \label{Fig.CDEWdetail}
\end{figure}

\begin{figure}\centering
\includegraphics[width=0.5\columnwidth]{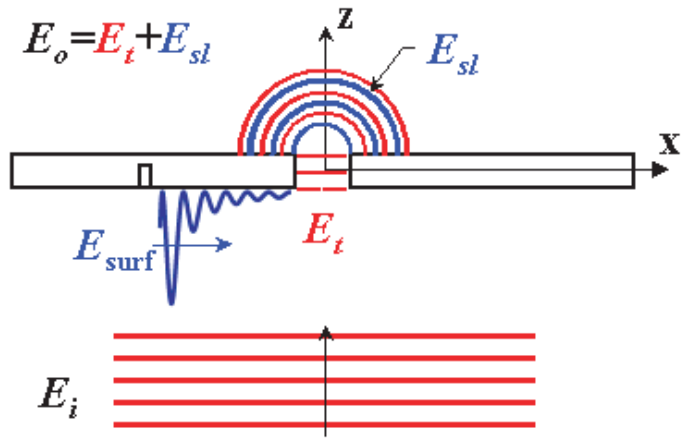}
\caption{The incoming plane wave $E_i$ impinges on the
subwavelength slit (or hole) and a groove milled on the input
side. The evanescent $E_{\mathrm{surf}}$ wave originates on the
surface at a slit-groove distance $x_{sg}$ and is indicated in
blue. In the model proposed in\,\cite{LT04} CDEWs travel along the
surface toward the slit where they reconvert to a propagating
field $E_{sl}$ and interfere with $E_t$, the propagating field
directly transmitted through the slit or hole.  The superposed
output field $E_o=E_t+E_{sl}$ propagates into the $z\geq 0$
half-space and the intensity of the interference figure
$I(\theta)$ is detected in the far
field.}\label{Fig:OnegrooveInput}
\end{figure}

The essential elements of the CDEW model can best be summarised
with reference to Fig.\,\ref{Fig.CDEWdetail}.
It is based on a solution to the 2-D Helmholtz equation in the
near field and subject to the slab-like boundary conditions of a
slit in an opaque screen. The basic expression describing the
scalar wave is
\begin{equation}
\left\lbrack{\nabla}^2 + k^2\right\rbrack E(x,z)=0
\end{equation} with $\nabla ^2={\partial}^2/\partial x^2
+{\partial}^2/\partial z^2$, $k=2\pi/\lambda$ and $E(x,z)$ the
amplitude of the wave propagating in the $x,z$ directions.
Kowarz\,\cite{K95} has written down the solution to this equation
for the case of an incident plane wave propagating in air ($n=1$)
with amplitude $E_i$ and propagation vector $k_0$ impinging on a
slit of width $d$ in an opaque screen. Specifying the coordinates
as shown in Fig.\,\ref{Fig.CDEWdetail}, the field solution
$E_{\mathrm{ev}}$ for the modes evanescent in $z$ at the $z=0$
boundary is
\begin{subequations}
\begin{eqnarray}
E_{\mathrm{ev}}(x,z=0)&=&-\frac{E_i}{\pi}\left\{
\mathrm{Si}\left\lbrack k\left(x+\frac{d}{2}\right)\right\rbrack -
\mathrm{Si}\left\lbrack k\left(x-\frac{d}{2}\right)\right\rbrack\right\}\quad\mbox{for}\quad |x|>d/2\label{Eq:Kowarz expression-a}\\
&=&\frac{E_i}{\pi}\left\{\pi- \mathrm{Si}\left\lbrack
k\left(x+\frac{d}{2}\right)\right\rbrack + \mathrm{Si}\left\lbrack
k\left(x-\frac{d}{2}\right)\right\rbrack\right\}\;\;\mbox{for}\;\;
|x|\leq
d/2\;\;\mbox{with}\;\;\mathrm{Si}(\alpha)\equiv\int_0^{\alpha}\frac{\sin
(t)}{t}dt\label{Eq:Kowarz expression-b}
\end{eqnarray}
\end{subequations}
The $k_z$ evanescent modes are determined by a
conservation-of-energy criterion,
\begin{equation}
k_z=\sqrt{k_0^2-k_x^2}\qquad k_x>k_0
\end{equation}
The form of the inhomogeneous or evanescent field on the $z=0$
boundary is shown in Fig.\,\ref{Fig.CDEWdetail}. At transverse
displacements from the slit $|x|>d/2$, the evanescent component of
the field at the surface $E_{\mathrm{ev}}(x,z=0)$ can be
represented to good approximation by the expression
\begin{equation}
E_{\mathrm{ev}}\simeq\frac{E_i}{\pi}\frac{d}{x}\cos
\left(k_{\mathrm{surf}}x+\pi/2\right)\label{Eq:CDEWcosrep}
\end{equation}
that describes a damped wave with amplitude decreasing as the
inverse of the distance from the launching edge of the slit, a
phase shift $\pi/2$ with respect to the propagating plane wave at
the midpoint of the slit and a wave vector
$k_{\mathrm{surf}}=2\pi/\lambda_{\mathrm{surf}}$. The wavelength
of the CDEW on the surface
$\lambda_{\mathrm{surf}}=\lambda_0/n_{\mathrm{surf}}$ where
$n_{\mathrm{surf}}$ is the surface index of refraction
(empirically, $n_{\mathrm{surf}}\simeq 1.04$). This surface wave
is actually a composite superposition of $k_x$ modes evanescent in
$z$, with $|k_x|>k_0$ and directed along the $\pm x$ axes.
\begin{equation}
E_{\mathrm{ev}}(x,z)= \frac{E_i}{\pi}\int_{\pm k_0}^{\pm\infty}
dk_x\frac{\sin (k_x\,
d/2)}{k_x}\exp(ik_x\,x)\exp(-k_zz)\label{Eq:IntegralofEvanesecentModes}
\end{equation}
Equation \ref{Eq:IntegralofEvanesecentModes} generalises the
expressions of Eqs.\,\ref{Eq:Kowarz expression-a},\,\ref{Eq:Kowarz
expression-b} to include the evanescent components above the $z=0$
plane. When the composite evanescent wave encounters a surface
discontinuity (a slit for example), a fraction of the surface wave
is reconverted to a distribution of ``homogeneous" or propagating
modes $|\mathbf{k}|=2\pi/\lambda_0$ at the site of the slit. In a
practical experiment, any real planar structure has two surfaces:
an ``input side" in the half-space $z<0$, containing the incoming
plane wave, and an ``output side" in the half-space $z\geq 0$,
containing the far-field propagating modes issuing from the output
surface and a photodetector. Experiments can be carried out by
fabricating subwavelength grooves on the input side, the output
side or both. The measurements reported here concern only the
input-side experiments (Fig.\,\ref{Fig:OnegrooveInput}).  Results
for output-side experiments will be reported later.

\section{Measurements and Results}
\begin{figure}\centering
\includegraphics[width=0.50\columnwidth]{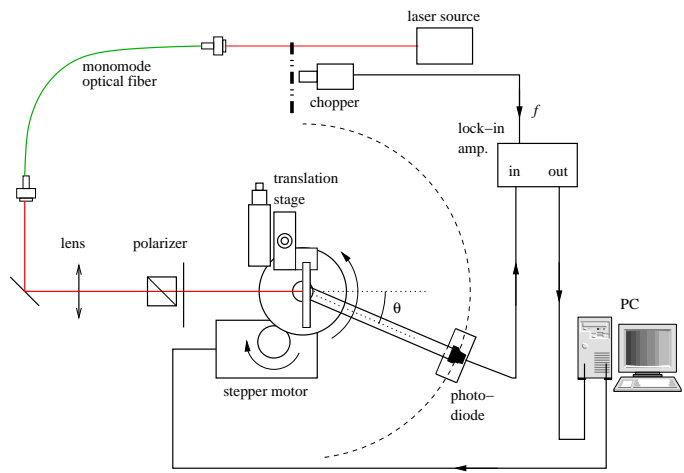}
\caption{Goniometer setup for measuring far-field light intensity
and angular distributions.  A stabilised single mode CW diode
laser, locked to a wavelength of 852 nm and modulated at 850 Hz by
a chopper wheel, is injected into a single-mode fibre and focused
onto the nanostructures mounted in a x-y translation stage as
shown. A stepper motor drives the goniometer arm, and the chopped
light intensity detected by the photodiode is fed to a lock-in
amplifier.  Output from the lock-in is registered by the PC that
also drives the stepper motor.  For the input-side experiments
described here the detector was always positioned at
$\theta=0^{\circ}$.}\label{Fig:goniometre}
\end{figure}

\begin{figure}\centering
\begin{minipage}[t]{0.45\columnwidth}\centering
\includegraphics*[width=0.8\columnwidth]{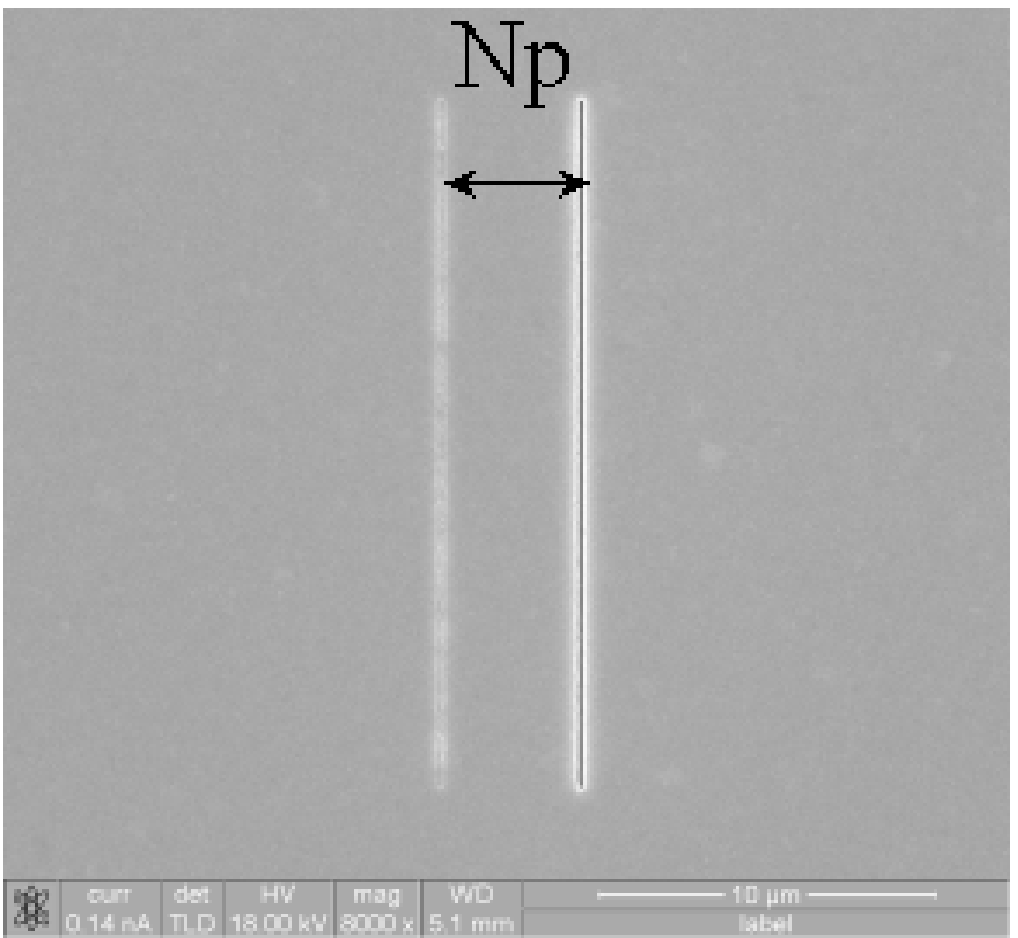}\caption{Scanning
electron microscope (SEM) image of one of the series of
single-slit, single-groove structures FIB milled into a 400 nm
thick silver layer deposited on flat quartz microscope slides 1 mm
thick. The width of both the slit and the groove is 100 nm, the
height $20\mu$m and the groove depth $\sim 100$ nm. The distance
$\mathrm{Np}$ is the pitch increment $\mathrm{p}=104\,$nm
multiplied by the number of increments
$\mathrm{N}$.}\label{Fig:OneSlitThinGroove}
\end{minipage}\hfill
\begin{minipage}[t]{0.45\columnwidth}\centering
\includegraphics*[width=0.8\columnwidth]{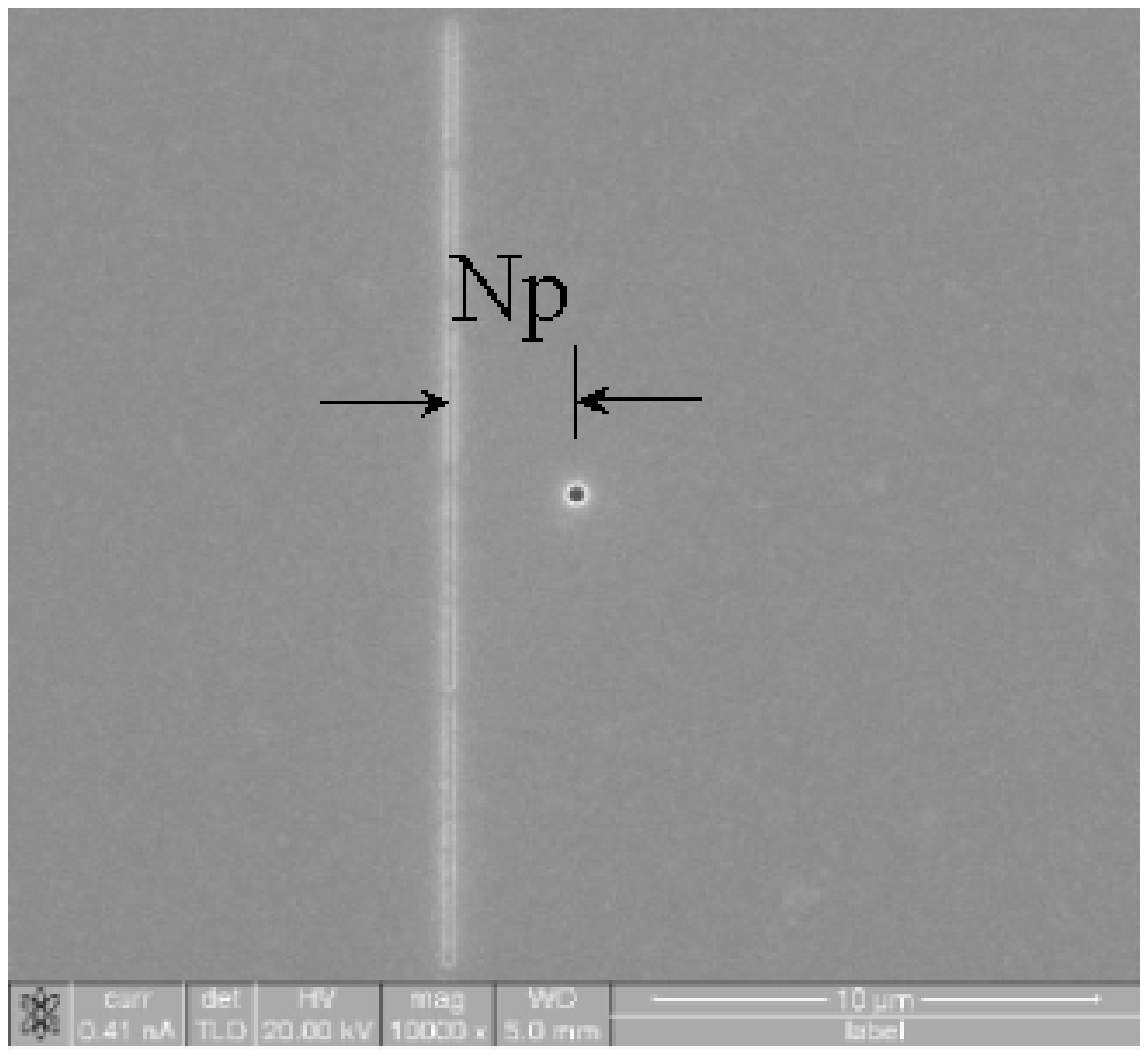}
\caption{SEM image of one of the series of single-groove,
single-hole structures fabricated similarly to the single-groove,
single-slit structures of Fig.\,\ref{Fig:OneSlitThinGroove}. The
silver layer for the groove-hole structures is 260 nm thick, the
width and depth of the groove is 100 nm and 70 nm respectively,
and the diameter of the hole is 300 nm. The distance $\mathrm{Np}$
is the pitch increment $\mathrm{p}=104\,$nm multiplied by the
number of increments $\mathrm{N}$}\label{Fig:SlitHole}
\end{minipage}
\end{figure}

\begin{figure}\centering
\begin{minipage}[t]{0.48\columnwidth}\centering
\includegraphics[width=\columnwidth]{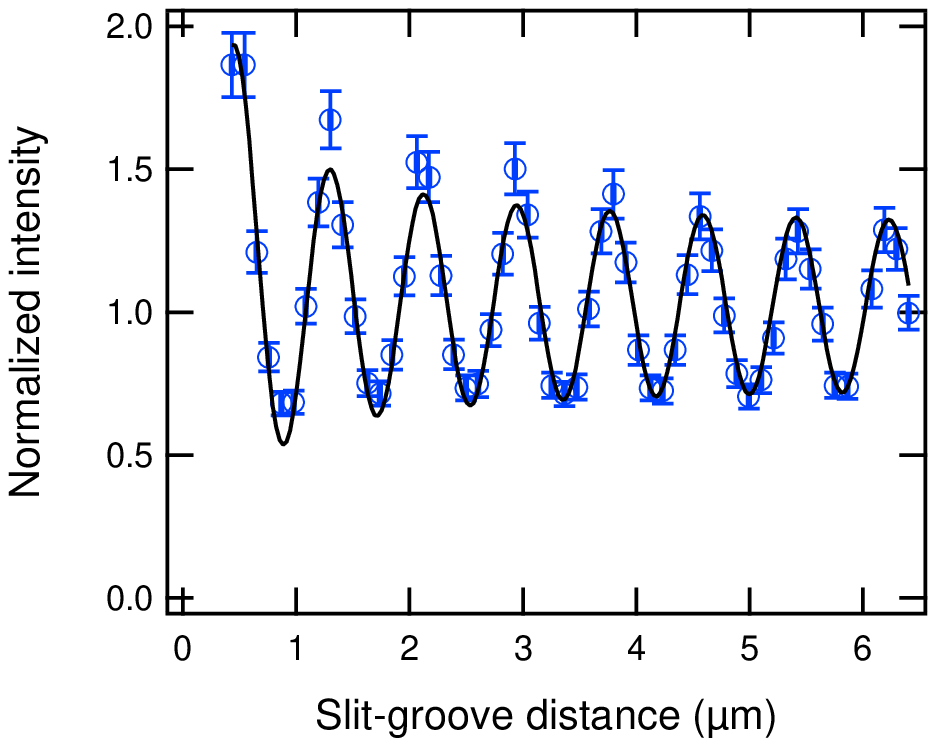}\caption{Normalised far-field
intensity $I/I_0$ as a function of slit-groove distance $x_{sg}$
for series of single-slit, single-groove structures mounted facing
the input side with respect to plane wave excitation.  Points are
the measured data through which the solid line,
Eq.\,\ref{Eq:EtaSlitInputSide}, is fitted with parameters
$\mu_{sl}=0.13\pm 0.01$, $\kappa_{sl}=0.12\pm 0.01\,\mu$m and
$\varphi_{\mathrm{int}}^{sl}=0.81\pm 0.02 \pi$. Error bars were
determined from variations in the measured intensities of the six
nominally identical naked slits (no flanking groove) used for
normalisation of each measurement.  Analysis of the frequency
spectrum of the fringe pattern for the slit-groove structures
results in the determination of a surface wavelength
$\lambda_{\mathrm{surf}}=819\pm 8\,$nm and an effective surface
index of refraction $n_{\mathrm{surf}}=1.04\pm 0.01$.}
\label{Fig:1stgInputSideResults}
\end{minipage}\hfill
\begin{minipage}[t]{0.48\columnwidth}\centering
\includegraphics[width=\columnwidth]{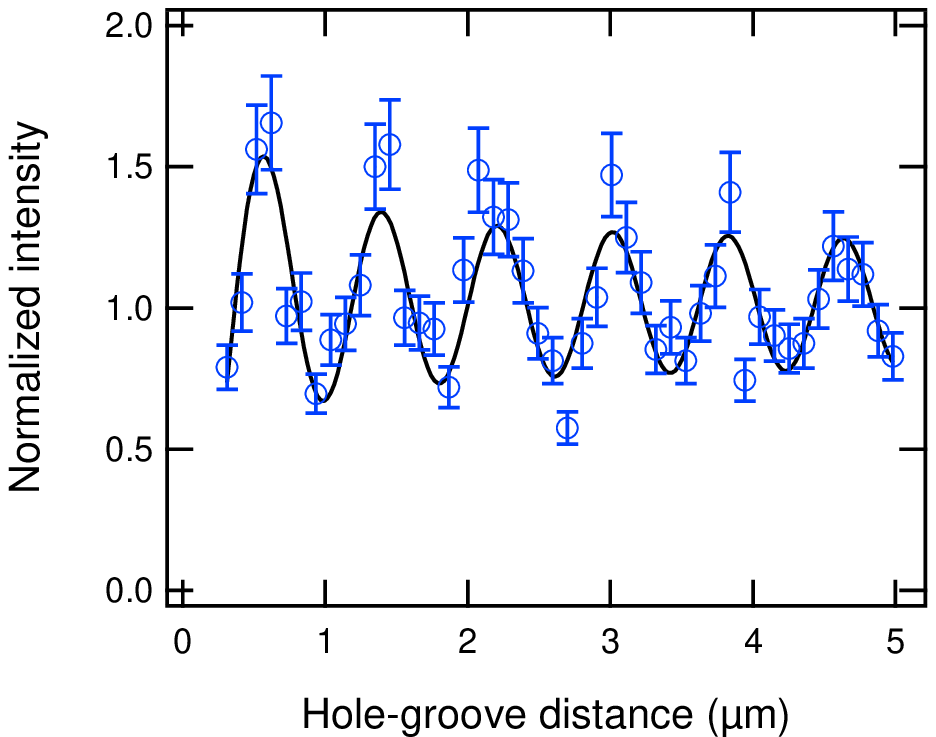}\caption{Normalised far-field intensity $I/I_0$ as a
function of hole-groove distance $x_{hg}$ for series of
single-hole, single-groove structures mounted facing the input
side with respect to plane-wave excitation.  Points are measured
data through which the solid line, Eq.\,\ref{Eq:EtaHoleInputSide},
is fitted with parameters $\mu_{hl}=0.10\pm 0.02$,
$\kappa_{hl}=0.08\pm 0.03\,\mu$m, and
$\varphi_{\mathrm{int}}^{hl}=0.55\pm 0.05\,\pi$.  Error bars were
determined from variations in the measured intensities of the six
nominally identical naked holes (no flanking groove) used for
normalisation of each measurement. Analysis of the frequency
spectrum of the fringe pattern for the hole-groove structures
results in the determination of a surface wavelength
$\lambda_{\mathrm{surf}}=811\pm 8\,$nm and an effective surface
index of refraction $n_{\mathrm{surf}}=1.05\pm
0.01$.}\label{Fig:hole-grooveResults}
\end{minipage}
\end{figure}

Measurements of the optical response of the slit-groove and
hole-groove structures were carried out using a home-built
goniometer shown in Fig.\,\ref{Fig:goniometre}, details of which
are described in the caption of Fig.\,\ref{Fig:goniometre} and in
the Methods section.
We have carried out a series of measurements on simple 1-D
structures to test the ``signature" predictions of the CDEW model,
\textit{viz.} (1) a composite surface wave expressed by
Eq.\,\ref{Eq:IntegralofEvanesecentModes} and approximately
represented by a damped wave, Eq.\,\ref{Eq:CDEWcosrep}; (2) a
phase shift of $\pi/2$ between the CDEW and the driving source
plane wave and (3) a wave amplitude that decreases inversely with
distance from the launching groove.  Figures
\ref{Fig:OneSlitThinGroove},\,\ref{Fig:SlitHole} show one of the
series of structures consisting of one slit and one groove and one
hole and one groove, respectively.  The slit-groove distance
$x_{sg}$ or hole-groove distance $x_{hg}$ is indicated as
$\mathrm{Np}$ where $\mathrm{p}$ is the basic unit of distance
increment, the ``pitch," and $\mathrm{N}$ is the number of
increments.  The pitch $\mathrm{p}$ was taken to be 104 nm,
approximately one-eighth the wavelength of the surface wave and
$\mathrm{N}$ was varied from 4 to 59. Structural details of these
devices are described in the captions of
\,Figs.\,\ref{Fig:OneSlitThinGroove},\,\ref{Fig:SlitHole} and in
the Methods section.
The slit(hole)-groove structures were mounted facing the input
side and exposed to plane-wave radiation from the focused
TEM$_{00}$ laser source. Measurements of light intensity on the
output side in the far field, 200 mm from the plane of the
structures, were carried out on the slit-groove structures using
the goniometer setup described in the Methods section.  The
results are shown in
Figs.\,\ref{Fig:1stgInputSideResults},\,\ref{Fig:hole-grooveResults}.
They show an oscillatory fringe pattern with amplitude damping out
to a distance of $\simeq 3-4\,\mu$m and maintaining an essentially
constant amplitude from that point out to the distance limit of
the measurements. As indicated in Fig.\,\ref{Fig:OnegrooveInput},
the fringe pattern results from interference between the mode
directly propagating through the slit (hole) at the input side
$E_t$ and a surface wave originating from the single-groove
structures $E_{\mathrm{surf}}$. The wave $E_{\mathrm{surf}}$ is
reconverted to a propagating mode at the slit or hole, and it is
this propagating mode that interferes with $E_t$. The frequency
and phase of the interference pattern is a function of the slit
(hole)-groove optical path and any intrinsic phase shift of the
surface wave itself. The normalised intensity $I/I_0$ of the
superposition term is given by
\begin{equation}
\frac{I}{I_0}=
1+\eta_i^2+2\eta_i\cos\gamma_i\quad\mbox{with}\quad\eta_i=\frac{\alpha\beta}{\delta}\label{Eq:normalised
intensity input side}
\end{equation}
where $\alpha=E_{\mathrm{surf}}/E_i$ is the fractional amplitude
of the surface wave launched from the incoming field $E_i$ at the
groove site, and $\beta$ is the further fraction of this surface
wave reconverted to a propagating wave in the slit, $E_{sl}=\beta
E_{\mathrm{surf}}=\beta\alpha E_i$. The fractional amplitude of
the directly transmitted component $E_t$ is $\delta$ and the phase
difference $\gamma_i$ between $E_t$ and $E_{sl}$ is the sum of two
terms,
\begin{equation}
\gamma_i=k_{\mathrm{surf}}x_{sg(hg)}+\varphi_{\mathrm{int}}
\label{Eq:ArgumentGamma}
\end{equation}The first term $k_{\mathrm{surf}}x_{sg(hg)}$ is the phase accumulated by the surface wave propagating from the
groove to the slit (hole) and the second term
$\varphi_{\mathrm{int}}$ is any phase shift intrinsic to the
surface wave. The term $\varphi_{\mathrm{int}}$ includes the
``signature" shift of the CDEW plus any phase shift associated
with the groove width and depth. Figures
\ref{Fig:1stgInputSideResults},\,\ref{Fig:hole-grooveResults}
present a direct measure of the normalised amplitude damping with
distance, $\eta_i=\eta_i(x)$ and the period and phase of the
oscillations, from which the wavelength $\lambda_{\mathrm{surf}}$
of the surface wave, the phase $\varphi_{\mathrm{int}}$, and the
effective surface index of refraction $n_{\mathrm{surf}}$ can be
determined. Analysis of the frequency spectrum of the fringe
pattern for the slit(hole) structures results in the determination
of a surface wavelength $\lambda_{\mathrm{surf}}=819(811)\pm
8\,$nm and an effective surface index of refraction
$n_{\mathrm{surf}}=1.04(1.05)\pm 0.01$. The amplitude $\eta_i$ of
the oscillatory term depends on the slit(hole)-groove distance,
and
Figs.\,\ref{Fig:1stgInputSideResults},\,\ref{Fig:hole-grooveResults}
show that $\eta_i$ falls of with increasing distance.  This
fall-off is fit to an expression with two terms: an inverse
distance dependence term plus a constant term.
\begin{subequations}
\begin{eqnarray}
\eta_{i}^{sl}(x_{sg})\cos (\gamma_i)&=&\left(\frac{\kappa_{sl}}{x_{sg}}+\mu_{sl}\right)\cos (k_{\mathrm{surf}}x_{sg}+\varphi_{\mathrm{int}}^{sl})\label{Eq:EtaSlitInputSide}\\
\eta_{i}^{hl}(x_{hg})\cos
(\gamma_i)&=&\left(\frac{\kappa_{hl}}{x_{hg}}+\mu_{hl}\right)\cos
(k_{\mathrm{surf}}x_{hg}+\varphi_{\mathrm{int}}^{hl})\label{Eq:EtaHoleInputSide}
\end{eqnarray}
\end{subequations}
The best-fit values for $\mu,\kappa,\varphi$ are indicated in the
captions of
Figs.\,\ref{Fig:1stgInputSideResults},\,\ref{Fig:hole-grooveResults}
and in Table \ref{Table:FitTable} for slit and hole structures,
respectively. The subscript $i$ and superscripts $sl,hl$ on $\eta$
refer to input-side, slit and hole measurements, respectively.

\section{Discussions and Conclusions}
The measured interference fringes on slit(hole) structures exhibit
the presence of a surface wave with wavelength $819(811)\pm 8$ nm
and therefore a surface index of refraction
$n_{\mathrm{surf}}=1.04(1.05)\pm 0.01$. The amplitude behavior of
these fringes is also similar.  Both the slit-groove and
hole-groove structures exhibit an initial amplitude fall-off with
increasing distance, damping to an essentially constant amplitude
at a distance $\simeq 6\,\mu$m. This behavior is fit to the
expressions in
Eqs.\,\ref{Eq:EtaSlitInputSide},\,\ref{Eq:EtaHoleInputSide}. The
fitting procedure is a linear regression varying the relative
contributions of the constant and decaying amplitude terms,
$\mu_{sl(hl)}$ and $\kappa_{sl(hl)}$ and the intrinsic phase shift
$\varphi_{\mathrm{int}}^{sl(hl)}$. The results are summarized in
Table \ref{Table:FitTable}.
\begin{table}
\caption{Fit parameters and measured $n_s$ for slit and hole
structures.  Error bars were determined from variations in the
measured intensities of the six nominally identical naked slits
and holes (no flanking groove) used for normalisation of each
measurement in
Figs.\,\ref{Fig:1stgInputSideResults},\,\ref{Fig:hole-grooveResults}.\label{Table:FitTable}}
\begin{ruledtabular}
\begin{tabular}{cccc}
parameter & slit structure & hole structure & SPP model\\\hline
$n_{\mathrm{surf}}$ & $1.04\pm 0.01$ & $1.05\pm 0.01$ & $1.015$\\
$\lambda_{\mathrm{surf}}$(nm) & $819\pm 8$ & $811\pm 8$ & $844$\\
$\mu$ & $0.13\pm 0.01$ & $0.13\pm 0.02$ \\
$\kappa $\,($\mu$m) & $0.12\pm 0.01$ & $0.020\pm 0.020$\\
$\varphi_{\mathrm{int}} (\pi)$ & $0.81\pm 0.02$ & $0.55\pm 0.05$
\end{tabular}
\end{ruledtabular}
\end{table}
\begin{table}
\caption{Surface plasmon parameters\label{Table:PlasmonParams}}
\begin{ruledtabular}
\begin{tabular}{ccccccc}
Reference&$\epsilon_{\mathrm{Ag}}^{\prime}$&$\epsilon_{\mathrm{Ag}}^{\prime\prime}$&$n_{\mathrm{sp}}$&L$_{\mathrm{abs}}$($\mu$m)&L$_{\mathrm{scat}}$($\mu$m)&L$_{\mathrm{rad}}$($\mu$m)\\[0.5ex]\hline
\footnote{Measurements on silver films used in these experiments
carried out at Caltech on a Sentech SE850 ellipsometer, 05
September 2005.}&-33.27&1.31&1.0154&109&$2.56\times
10^4$&$5.00\times
10^4${\rule[-1mm]{0mm}{5mm}}\\
\cite{JC75}&-34.&0.46&1.015&326&$2.6\times 10^4$&$5.0\times 10^4$\\
\cite{P85}&-32.4&1.74&1.0158&78.0&$2.51\times 10^4$&$2.68\times
10^4$
\end{tabular}
\end{ruledtabular}
\end{table}

How do these results compare to CDEW or SPP models? In the CDEW
picture, the groove launches a surface wave on the input side of
the silver film that is detected by interference with the directly
transmitted wave through the hole or slit, in the far field, on
the output side of the structure. The amplitude of this surface
wave is predicted to damp as the inverse distance between the
groove and the slit or hole.  Figures
\ref{Fig:1stgInputSideResults},\,\ref{Fig:hole-grooveResults} show
an initial decrease in amplitude with increasing distance out to
about $3-4\,\mu$m, but that the amplitude thereafter remains
essentially constant.   The solid curves in
Figs.\,\ref{Fig:1stgInputSideResults},\,\ref{Fig:hole-grooveResults}
fit this amplitude decrease to an inverse distance dependence
(Eqs.\,\ref{Eq:EtaSlitInputSide},\,\ref{Eq:EtaHoleInputSide}).
However, the damping might also plausibly fit an exponential
decrease which would be expected from surface plasmon dissipative
processes such as absorption by the silver film or scattering due
to surface roughness.  In order to check this possibility we have
measured the properties of the silver films used in these studies.
Table \ref{Table:PlasmonParams} summarises these properties and
compares them to previously reported
measurements\,\cite{JC75,P85}.  The dielectric constant at 852\,nm
$\epsilon_{\mathrm{Ag}}=\epsilon^{\prime}_{\mathrm{Ag}}+\epsilon^{\prime\prime}_{\mathrm{Ag}}$
was measured by ellipsometery and surface roughness parameters
determined by atomic force microscopy (AFM). The root-mean-square
(rms) height of the films was measured to be $\delta=1.29\,$nm and
the correlation length $\sigma=154.3\,$nm. From the imaginary term
of the dielectric constant $\epsilon^{\prime\prime}_{\mathrm{Ag}}$
and the parameters $\delta, \sigma$ the expected propagation
lengths of surface plasmons against absorption, surface
scattering, and reradiation, $L_{\mathrm{ams}}, L_{\mathrm{scat}},
L_{\mathrm{rad}}$, can be calculated\,\cite{Raether88}. It is
clear from columns 5-7 of Table\,\ref{Table:PlasmonParams} that
these loss processes cannot account for the observed damping
within $3\,\mu$m of hole-groove distance.

The constant amplitude beyond $\sim 3-4\,\mu$m is consistent with
a persistant surface wave.  Indeed we have recorded measurements
(not presented here) of the surface wave persisting at least to
$\simeq 30\,\mu$m slit-groove distance.  It is important to
emphasise, however, that $\lambda_{\mathrm{surf}}$ and
$n_{\mathrm{surf}}$ deviate significantly for those expected for a
pure SPP on a plane silver surface.  Interferometry measurements
of the surface waves on ``output side" slit-groove structures (not
reported here) confirm the value of $n_{\mathrm{surf}}$ in Table
\ref{Table:FitTable}, and we believe that conventional,
infinite-plane SPP theory \cite{Raether88} is not adequate to
explain these results. We note that persistent surface waves over
$\sim 10\,\mu$m distances have also been reported in a double slit
experiment\,\cite{SKD05} and interpreted as SPPs\,\cite{LHR05}.

As indicated in Table \ref{Table:FitTable}, intrinsic phase for
the slit-groove and hole-groove structures respectively are
$\varphi_{\mathrm{int}}^{sl}=0.81\,\pi$ and
$\varphi_{\mathrm{int}}^{hl}=0.55\,\pi$. Although one contribution
to these phase shifts may be the CDEW ``signature" phase shift of
$\pi/2$, it is known from earlier studies that the specific form
(width and depth) of the grooves themselves, can introduce phase
shifts into the scattered wave\,\cite{VLE03}. We have determined
the nature of these groove-induced phase shifts and resonances by
measuring interference fringes arising from surface waves launched
on the ``output-side" of slit-groove structures. These results,
that will be reported in a subsequent publication, support the
existence of an intrinsic phase shift close to $\pi/2$.

The interpretation that emerges from these results is that the
subwavelength groove originates persistant, long-range surface
waves by a two-step process: (1) the incoming TM polarised plane
wave scatters from the groove and generates in its immediate
vicinity on the surface a broad, CDEW-like distribution of
diffracted evanescent waves, and (2) this broad-band local surface
``emitter" excites, within a distance of  $\simeq 3-4\,\mu$m, a
long-range surface wave response. The near-term rapid amplitude
decrease in the interference fringes of
Figs.\,\ref{Fig:1stgInputSideResults},\,\ref{Fig:hole-grooveResults}
is evidence of this evanescent surface wave diffraction very near
the groove. Persistant amplitude out to tens of microns is
evidence for some kind of surface wave guided mode. It is
significant to note that the wavelength and phase of the
interference fringes do not shift over the entire range of the
measurements.  The initial diffracted surface wave components
extend over a broad range of evanescent modes, $k_x>k_0$,
including the conventional $k_{\mathrm{SPP}}$. Therefore it is to
be expected that the local surface wave emitter excites this
surface mode.  We emphasise, however, that our measurements show
that the wavelength of this persistant wave does not correspond to
$k_{\mathrm{SPP}}$ and that, when the phase lag associated the
groove itself is taken into account,  the intrinsic phase of the
surface wave with respect to the directly transmitted wave is
close to $\pi/2$. The disaccord between $\lambda_{\mathrm{SPP}}$
and $\lambda_{\mathrm{surf}}$ is for the present a matter of
speculation.  Perhaps plasmon ``leaky waves"\,\cite{Petit80} that
transport energy very slowly away from the surface contribute to
the spectrum of long-range surface excitation resulting in an
effective wavelength shift; or perhaps, despite our ellipsometry
measurements, the surface index of refraction of the metal film is
slightly modified by some uncontrolled chemical or material
process\,\cite{MM00}. At a more practical level, these results
indicate that it might be much easier to couple to surface guided
waves than was previously thought. Conventional wisdom asserts
that because the SPP lies to the right of the ``light line" on the
metal surface dispersion curve, a grating or prism is needed to
achieve efficient optical coupling \,\cite{Raether88}. A simple
abrupt discontinuity in the surface, such as a slit or groove,
appears to serve as an efficient coupler.  Further studies are
needed to understand the properties of the generated long-range
persistant wave and to optimise the efficiency of this
groove-coupling process.
\section{Methods}
\subsection{Structure fabrication}
The structures consist of a single subwavelength slit or hole
flanked by one subwavelength groove. The grooves have a width of
100 nm and a nominal depth of 100 nm for the slit-groove
structures and 70 nm for the hole-groove structures.  The
slit-groove distance ($x_{sg}$) or hole-groove distance ($x_{hg}$)
is systematically incremented in the fabrication process.  The
subwavelength structures are fabricated by focused ion beam (FIB)
milling (FEI Nova-600 Dual-Beam system, Ga$^+$ ions, 30keV) into a
layer of silver evaporated onto flat fused silica microscope
slides.  A low beam current (50\,pA) was used in order to achieve
surface features defined with a lateral precision on the order of
10\,nm and characterised by near-vertical sidewalls and a minimal
amount of edge rounding. Since it enables delivery of a variable
ion dose to each pixel of the writing field, FIB milling
conveniently allows the multiple-depth topography characteristic
of the present devices to be formed in a single, self-aligned
step. A 2-D matrix of structures is milled into the silver layer.
Each matrix consists of 63 structures, nine columns, separated by
1.5 mm, and seven rows, separated by 2 mm. The first column
contains only the slit with no flanking grooves. Light
transmission through the slits in this column is used to normalise
the transmission in the remaining columns. Variations in
transmission through each of the elements in the ``slits only"
column provide a measure of the uniformity of the FIB fabrication
process.  Each entire matrix of structures is flanked on one side
by a small round hole and on the other by a line grating for
absolute reference positioning and angular alignment of the
structure matrix with respect to the input laser beam. The square
microscope slides themselves, commercially available from SPI
Supplies, are 25 mm on a side and 1 mm thick.
\subsection{Measurement Setup}
Details of the experimental setup are as follows.
 Output from a diode laser source, temperature stabilised and
frequency-locked to $^2\mathrm{S}_{1/2}(\mathrm{F}=4)\rightarrow$
$ ^2$P$_{3/2}(\mathrm{F}=4,5)$ crossover feature in a Cs saturated
absorption cell, is modulated at 850 Hz by a mechanical chopper,
fed to a monomode optical fibre, focused and finally linearly
polarised before impinging on the subwavelength structure mounted
in the sample holder.  The beam waist diameter and confocal
parameter of the illuminating source are 300 $\mu$m and 33 cm,
respectively. Throughout this series of measurements the laser
power density was maintained $\sim 1$Wcm$^{-2}$. The sample holder
itself is fixed to a precision x-y translator, and multiple
structures, FIB-milled in a 2-D array on a single substrate, are
successively positioned at the laser beam waist.  The optical
response of the structures is synchronously detected by a
photodiode and registered on a laboratory computer as indicated in
Fig.\,\ref{Fig:goniometre}.

\begin{acknowledgments}
Support from the Minist{\`e}re d{\'e}l{\'e}gu{\'e} {\`a} l'Enseignement sup{\'e}rieur et {\`a}
la Recherche under the programme
ACI-``Nanosciences-Nanotechnologies," the R{\'e}gion Midi-Pyr{\'e}n{\'e}es
[SFC/CR 02/22], and FASTNet [HPRN-CT-2002-00304]\,EU Research
Training Network, is gratefully acknowledged as is support from
the Caltech Kavli Nanoscience Institute and from the AFOSR under
Plasmon MURI FA9550-04-1-0434.  Discussions and technical
assistance from P. Lalanne, R. Mathevet, F. Kalkum, G. Derose, A.
Scherer, D. Pacifici, J. Dionne, R. Walters and H. Atwater are
also gratefully acknowledged.
\end{acknowledgments}

\end{document}